
\voffset=1.5truein
\hoffset=1truein
\hsize=4.5truein
\vsize=7.125truein
\parindent=7mm
\parskip=5pt plus 5pt
\baselineskip=1.05\baselineskip

\font\smallrm=cmr9
\font\eighttt=cmtt8
\font\smallit=cmti9
\font\sevenit=cmti7
\font\eigit=cmti8
\font\ninerm=cmr9 \font\ninesy=cmsy9
\font\ninei=cmmi9
\font\nineit=cmti9 \font\ninebf=cmbx9
\font\ninesl=cmsl9
\def\ninepoint{\textfont0=\ninerm\textfont1=\ninei
  \textfont2=\ninesy\scriptfont0=\sevenrm\scriptfont1=\seveni
  \scriptfont2=\sevensy\scriptscriptfont0=\fiverm\scriptscriptfont1=\fivei
  \scriptscriptfont2=\fivesy\def\rm{\fam0\ninerm}\let\it=\nineit\let\bf=\ninebf
  \let\sl=\ninesl\baselineskip=12pt\rm}

\font\eightrm=cmr8
\font\eighti=cmmi8
\font\eightsl=cmsl8
\def\eightpoint{\textfont0=\eightrm\textfont1=\eighti
  \textfont2=\ninesy\scriptfont0=\sevenrm\scriptfont1=\seveni
  \scriptfont2=\sevensy\scriptscriptfont0=\fiverm\scriptscriptfont1=\fivei
  \scriptscriptfont2=\fivesy\def\rm{\fam0\ninerm}\let\it=\nineit\let\bf=\ninebf
  \let\sl=\eightsl\baselineskip=9pt\rm}

\mathchardef\Gammait"0100
\mathchardef\Deltait"0101
\mathchardef\Phiit"0108
\scriptfont\itfam=\sevenit
\newfam\srmfam
\textfont\srmfam=\smallrm

\catcode`\?=11
\def\X?magic{^^C}
\def\mat?#1{$#1$}
\def\X?#1 $#2$#3\endX?{\def\X?test{#2}\ifx\X?test\X?magic
   \def\X?next{#1}\else\def\X?next{\X?#1\/ \mat?{#2}#3\endX?}\fi\X?next}
\long\outer\def\icproclaim#1. #2\par{\medbreak
    \noindent{\bf#1.\enspace}{\it\X?#2 $^^C$\endX?}\par
    \ifdim\lastskip<\medskipamount\removelastskip\penalty55\medskip\fi}
\catcode`\?=12

\def\blackslug{\hbox{\kern1pt\vrule height6pt width4pt depth1pt\kern1pt}}
\def\proof{\par\noindent{\bf Proof.\enspace}\rm}
\def\eop{\penalty 500\hbox{\quad\blackslug}\ifmmode\else\par
    \vskip4.5pt plus3pt minus2pt\fi}

\outer\def\beginsection#1\par{\goodbreak\bigskip\vskip\parskip
  \message{#1}\leftline{\bf#1}\nobreak\smallskip\indent}
\def\({\bigl(}
\def\){\bigr)}

\def\frac#1#2{{\textstyle{#1\over#2}}}
\fontdimen16\tensy=1\fontdimen17\tensy

\def\0{{\eighttt0}\kern0.51mm}
\def\1{{\eighttt1}\kern0.51mm}

\def\R{{\cal R}}

\def\VG{{\it VG}}
\def\EG{{\it EG}}

\font\nineasc=cmtt9
\def\uncatcodespecials{\def\do##1{\catcode`##1=12 }\dospecials}
\def\listing#1 {\par\begingroup\setupverbatim\input#1 \endgroup}
\def\setupverbatim{\nineasc\def\par{\leavevmode\endgraf}\parskip=0pt plus 1pt
\parindent=0pt \baselineskip=4truemm \obeylines\uncatcodespecials\obeyspaces
\def~{\char"7E }\catcode`\~=\active}
{\obeyspaces\global\let =\ }


\centerline{\bf NOTE}

\bigskip
\centerline{\bf Computation of the Ramsey Number $R(W_5,K_5)$}

{\bigskip
\baselineskip=0.85\baselineskip
\smallit\tabskip=10pt plus 50pt

$$\halign to 0.97\hsize{\hfil#\hfil&\hfil#\hfil\cr
\rm Stanis\l aw P. Radziszowski&\rm Kung-Kuen Tse\cr
\eigit and \rm Joshua Stinehour&Department of Computer\cr
Department of Computer Science&Science and Mathematics\cr
Rochester Institute of Technology&Kean University\cr
Rochester, NY 14623&Union, NJ 07083\cr
\nineasc $\{$spr,jjs9804$\}$@cs.rit.edu&\nineasc ktse@kean.edu\cr}$$

\bigskip}


\bigskip
\bigskip
\bigskip

\centerline{\bf Abstract}

{\parindent=0.3truein\narrower\narrower\narrower\narrower\narrower
We determine the value of the Ramsey number $R(W_5,K_5)$ to be 27,
where $W_5 = K_1 + C_4$ is the 4-spoked wheel of order 5. This solves
one of the four remaining open cases in the tables given in 1989
by George R. T. Hendry, which included the Ramsey numbers $R(G,H)$
for all pairs of graphs $G$ and $H$ having five vertices, except
seven entries. In addition, we show that there exists a unique
up to isomorphism critical Ramsey graph for $W_5$ versus $K_5$.
Our results are based on computer algorithms.}

\bigskip
\bigskip
\noindent
{\bf Keywords:} Ramsey numbers, graph algorithms\hfill\break
{\bf AMS subject classification:} 05C55


\bigskip
\bigskip
\beginsection 1. Overview

This note is a continuation of the work reported in [1], which
contained the result $R(B_3,K_5)=20$. It has similar origins, and
also the scenario of work arrangements was similar. The main result
of this note, the equality $R(W_5,K_5)=27$, was obtained with the help
of computer algorithms that were a part of an MS thesis work
by Josh Stinehour, under supervision of Stanis\l aw Radziszowski,
and verified with independently written programs by Kung-Kuen Tse.
We will use the same definitions and notation as in [1], which
appeared in this Bulletin [Vol. 41 (2004) 71-76].

In 1989, George R. T. Hendry [2] presented a table of Ramsey numbers
$R(G,H)$ for all pairs of graphs $G$ and $H$ having five vertices, with
the exception of seven cases: $R(C_5+e,K_5)$, $R(W_5,K_5-e)$,
$R(B_3,K_5)$, $R(W_5,K_5)$, $R(K_5-P_3,K_5)$, $R(K_5-e,K_5)$ and
$R(K_5,K_5)$.  Until now, only three of these open cases have been 
solved: $R(C_5+e,K_5)=17$, $R(W_5,K_5-e)=17$ and $R(B_3,K_5)=20$.
A regularly updated survey by the first author [5] reports on old and the
most recent results on various types of Ramsey numbers, including those of
the form $R(G,H)$.  In particular, [5] lists the developments related to
all seven cases missing in Hendry's table, and gives references to
papers discussing them.

In this work, we eliminate one of these open cases by computing
$R(W_5,K_5)=27$. This result improves the bounds
$27 \le R(W_5,K_5) \le 29$ given in [2].
In addition, we show that there exists a unique
up to isomorphism critical graph, i.e. $|\R(W_5,K_5;26)|=1$.

Thus, the remaining open cases of two-color Ramsey numbers
for general graphs on at most five vertices are:
$25 \le R(K_5-P_3,K_5) \le 28$,
$30 \le R(K_5-e,K_5) \le 34$, and $43 \le R(K_5,K_5) \le 49$
(see [5]  for references to all bounds). The expected
difficulty of these cases is discussed in [1].


\bigskip

\beginsection 2. Enumerations and Results

It is known that $R(C_4,K_5)=14$ and $R(W_5,K_4)=17$ (see [5]).
The set of all 1888 graphs in $\R(C_4,K_5)$ was enumerated in [6],
and for this project fairly simple algorithms were sufficient
to generate all 3071561 graphs in $\R(W_5,K_4)$. The statistics
of both families by the number of graphs with fixed number of vertices
are given in Table I.

\medskip
\midinsert
$$\vbox{\tabskip=0pt
\eightpoint
\halign to 0.47\hsize{
\hfil#\tabskip=10pt plus50pt
  &\hfil#&\hfil#&\hfil#&\hfil#&\tabskip=10pt plus50pt \hfil#\cr
\noalign{\kern1mm\hrule\kern1mm}
$s$&$|\R(C_4,K_5;s)|$&$|\R(W_5,K_4;s)|$\cr
\noalign{\kern1mm\hrule\kern1mm}
1&1&1\cr
2&2&2\cr
3&4&4\cr
4&8&10\cr
5&17&26\cr
6&38&94\cr
7&85&401\cr
8&190&2307\cr
9&385&15452\cr
10&574&104314\cr
11&457&531892\cr
12&126&1437877\cr
13&1&865055\cr
14&&111153\cr
15&&2891\cr
16&&82\cr
\noalign{\kern1mm\hrule\kern1mm}
total&1888&3071561\cr
\noalign{\kern1mm\hrule\kern1mm}
}}$$

\centerline{\bf Table I. \rm Statistics for $\R(C_4,K_5)$ and $\R(W_5,K_4)$.}
\kern2mm
\endinsert

For a graph $G$,
if $v\in\VG$, then $N_G(v) = \{ w \in \VG \,|\, vw\in \EG \}$.
The subgraph of $G$ induced by $W$ will be denoted by $G[W]$.
Also, for $v \in \VG$, define the induced subgraphs $G_v^+ =G[N_G(v)]$
and $G_v^- =G[\VG-N_G(v)-\{v\}]$.
Note that if $G \in \R(W_5,K_5;n)$ and $v \in \VG$,
then necessarily $G_v^+ \in \R(C_4,K_5;d)$, where
$d=\deg_G(v)$, and $G_v^- \in \R(W_5,K_4;n-d-1)$.

For all cases, the construction of $\R(W_5,K_5;n)$
proceeds by using the results in Table I and applying
the gluing algorithm to
$G_v^+ \in \R(C_4,K_5;s)$ and $G_v^- \in \R(W_5,K_4;t)$,
for all possible $s$ and $t$ satisfying $s+t+1=n$.
The gluing algorithm used in this work was similar to that
described in [1, 4, 6], except for some modifications
which were needed in order to
avoid the graph $W_5$ instead of $B_3$, $K_4$ or $C_4$. 

All $(W_5,K_5;26)$-graphs were obtained by performing gluing of graphs
$G_v^+$ to $G_v^-$ for $s \in \{9, 10, 11, 12, 13\}$ and $t=25-s$.
Table II presents the statistics of the gluings that were completed.
The computations led to the unique $(W_5,K_5;26)$-graph,
which is cyclic and regular of degree 9, with
the edges connecting pairs of vertices belonging to ${\cal Z}_{26}$
in circular distances 1, 5, 8, 12 and 13.

\bigskip
\midinsert
$$\vbox{\tabskip=0pt
\eightpoint
\halign to 0.77\hsize{
\hfil#\tabskip=10pt plus50pt
  &\hfil#&\hfil#&\hfil#&\hfil#&\hfil#&\tabskip=10pt plus50pt \hfil#\cr
\noalign{\kern1mm\hrule\kern1mm}
$s$&$|\R(C_4,K_5;s)|$&$|\R(W_5,K_4;25-s)|$&no. of generated&\cr
   &                 &                    &$(W_5,K_5;26)$-graphs\cr
\noalign{\kern1mm\hrule\kern1mm}
9&385&82&1\cr
10&574&2891&0\cr
11&457&111153&0\cr
12&126&865055&0\cr
13&1&1437877&0\cr
\noalign{\kern1mm\hrule\kern1mm}
}}$$

\centerline{\bf Table II. \rm Statistics for computation of $(W_5,K_5;26)$-graphs}
\kern2mm
\endinsert

All $(W_5,K_5;27)$-graphs were obtained in two ways: by performing gluing
as above for $s \in \{10, 11, 12, 13\}$, $t=26-s$, and independently
by constructing and $(W_5,K_5)$-filtering all one-vertex
extensions of the unique $(W_5,K_5;26)$-graph. Both paths led to no graphs,
and thus $R(W_5,K_5)=27$.

\bigskip
\noindent
{\bf{Theorem.}} $R(W_5,K_5) = 27$.

\proof
The computations and results described above prove that no\break
$(W_5,K_5;27)$-graph exists, so $R(W_5,K_5) \le 27$.
It is easy to verify that a cyclic graph with the edges joining
vertices belonging to ${\cal Z}_{26}$ which are
in circular distance 1, 5, 8, 12 or 13, has no $W_5$ and no
$\overline{K}_5$. This implies the lower bound.
\eop

\bigskip

Two separate implementations of the algorithms were prepared and their
results compared. In order to corroborate the correctness of both
implementations, we have performed a number of gluings yielding
large output. Table III lists some special cases of gluing
instances producing a large number of graphs in $\R(W_5,K_5)$ on which
the two implementations agreed exactly.
The computational effort of this project was moderate ---
all computations can now be repeated overnight on a
small local departmental network.

\bigskip
\midinsert
$$\vbox{\tabskip=0pt
\eightpoint
\halign to 0.81\hsize{
\hfil#\tabskip=10pt plus50pt
  &\hfil#&\hfil#&\hfil#&\hfil#&\hfil#&\hfil#&\tabskip=10pt plus50pt \hfil#\cr
\noalign{\kern1mm\hrule\kern1mm}
   & & & & no. of generated graphs\cr
$s$&$|\R(C_4,K_5;s)|$&$t$&$|\R(W_5,K_4;t)|$&$G$ with $\delta(G)=s$ in\cr
   & & & & $\R(W_5,K_5;s+t+1)$\cr
\noalign{\kern1mm\hrule\kern1mm}
6&38&16&82&869853\cr
7&85&16&82&17421\cr
8&190&15&2891&1768\cr
\noalign{\kern1mm\hrule\kern1mm}
}}$$

\centerline{\bf Table III. \rm Further data on generated $(W_5,K_5)$-graphs}
\kern2mm
\endinsert

A general utility program for graph isomorph rejection, {\it nauty} [3],
together with other graph manipulation tools,
written by Brendan McKay, was used extensively.
All graphs whose statistics were given in this paper
are available from the authors.



\bigskip
\vskip 0pt plus 100pt
\goodbreak

\beginsection  References

\vskip 0pt plus -100pt
\vskip -2mm

\baselineskip=0.95\baselineskip
\parskip=2mm minus 2mm
\frenchspacing
\def\ref[#1]{\item{[#1]}}
\parindent=1.3\parindent

\ref[1] A. Babak, S. P. Radziszowski and Kung-Kuen Tse,
   Computation of the Ramsey Number $R(B_3,K_5)$,
   {\it Bulletin of the Institute of Combinatorics and Its Applications},
   {\bf 41} (2004) 71-76.

\ref[2] G. R. T. Hendry,
   Ramsey Numbers for Graphs with Five Vertices,
   {\it Journal of Graph Theory},
   {\bf 13} (1989) 245--248.

\ref[3] B. D. McKay, nauty users' guide (version 1.5), Technical
   Report TR-CS-90-02, Computer Science Department, Australian
   National University, 1990. Source code available at\hfill\break
   {\tt http://cs.anu.edu.au/people/bdm/nauty}.

\ref[4] B. D. McKay and S. P. Radziszowski, $R(4,5)=25$,
   {\it Journal of Graph Theory}, {\bf 19} (1995) 309--322.

\ref[5] S. P. Radziszowski, Small Ramsey Numbers,
   {\it Electronic Journal of Combinatorics}, Dynamic Survey~1,
   revision \#10, 48 pages, July 2004,
   {\tt http://www.combinatorics.org/}.

\ref[6] S. P. Radziszowski and Kung-Kuen Tse,
A Computational Approach for the Ramsey Numbers $R(C_4,K_n)$,
{\it Journal of Combinatorial Mathematics and Combinatorial Computing},
{\bf 42} (2002) 195--207.

\bye